\documentclass{ws-procs9x6}
\newcommand{\lwig}{\mbox{\,\raisebox{.3ex}
    {$<$}$\!\!\!\!\!$\raisebox{-.9ex}{$\sim$}\,}}
\newcommand{\gwig}{\mbox{\,\raisebox{.3ex}
    {$>$}$\!\!\!\!\!$\raisebox{-.9ex}{$\sim$}}\,}
\newcommand{\rav}{\langle\rho\rangle}
\newcommand{\xbj}{x_{\rm Bj}}

\newcommand{\vl}{\mathbf{l}}
\newcommand{\vr}{\mathbf{r}}

\newcommand{\vb}{\mathbf{b}}
\newcommand{\vx}{\mathbf{x}}
\newcommand{\sidp}{\sigma_\text{\tiny DP}}

\def\Journal#1#2#3#4{{#1}{\bf #2}, #3 (#4)}
\def\NPA{{\em Nucl. Phys.} {\bf A}}
\def\NPB{{\em Nucl. Phys.} {\bf B}}
\def\PLB{{\em Phys. Lett.}  {\bf B}}
\def\PRL{\em Phys. Rev. Lett. }
\def\PRD{{\em Phys. Rev.} {\bf D}}

\def\ZPC{{\em Z. Phys.} {\bf C}}
\def\EPJC{{\em Eur. Phys. J.} {\bf C}} 

\def\CPC{\em Comput. Phys. Commun. }

\def\JPG{{\em J. Phys.} {\bf G}}

\def\APPB{{\em Acta Phys.\ Polon.} {\bf B}}
\def\APH{\em Annals Phys. } 
\def\SNP{\em Sov. J. Nucl. Phys. }
\def\JETP{\em Sov. Phys. JETP }
\def\PRP{\em Phys. Rept. }
\begin{document}
\title{
\vspace{-3cm}
{\normalsize\rightline{DESY 03-217}\rightline{\lowercase{hep-ph/0401137}}}
\vskip 1cm
Instanton-Driven Saturation at Small $x$\footnote{
\uppercase{P}resented at the \uppercase{R}ingberg
\uppercase{W}orkshop: {\it \uppercase{N}ew \uppercase{T}rends in
\uppercase{HERA}  \uppercase{P}hysics 2003}, \uppercase{S}ep. 28
-- \uppercase{O}ct. 3, 2003,  
\uppercase{R}ingberg \uppercase{C}astle,
\uppercase{R}ottach-\uppercase{E}gern, \uppercase{G}ermany.}}

\author{F. SCHREMPP} 

\address{Deutsches Elektronen Synchrotron (DESY), \\
Notkestrasse 85, \\ 
D-22607 Hamburg, Germany\\ 
E-mail: fridger.schrempp@desy.de}

\author{A. UTERMANN}
\address{Institut  f{\"u}r Theoretische Physik der
Universit{\"a}t Heidelberg,\\
Philosophenweg 16, \\
D-69120 Heidelberg, Germany\\
E-mail: A.Utermann@thphys.uni-heidelberg.de
}

%%%%%%%%%%%%%%%%%%%%%%%%%%%%%%%%%%%%%%%%%%%%%%%%%%%%%%%%%%%%%%
% You may repeat \author \address as often as necessary      %
%%%%%%%%%%%%%%%%%%%%%%%%%%%%%%%%%%%%%%%%%%%%%%%%%%%%%%%%%%%%%%

\maketitle

\abstracts{We report on the interesting 
possibility of instanton-driven gluon saturation in lepton-nucleon
scattering at small Bjorken-$x$. Our results crucially involve
non-perturbative information from high-quality lattice
simulations. The conspicuous, intrinsic 
instanton size scale $\rav \approx 0.5$ fm, as known from the lattice,
turns out to determine the saturation scale. A central result is
the identification of the ``colour glass condensate'' with the
QCD-sphaleron state.}  

%\section{Setting the stage}
\section{Motivation}
\subsection{Saturation in the Parton Picture}
One of the most important observations from HERA is the strong rise of
the gluon distribution at small Bjorken-$x$\,\cite{HERA}. On the one
hand, this rise is predicted by the DGLAP evolution
equations\cite{DGLAP} at high $Q^2$ and thus supports QCD\cite{riseDGLAP}. On
the other hand, an undamped rise will eventually violate 
unitarity. The reason for the latter problem is known to be
buried in the linear nature of the DGLAP- and the
BFKL-equations\cite{BFKL}: For decreasing Bjorken-$x$, the number of
partons in the 
proton rises, while their effective size $\sim 1/Q$ increases with
decreasing $Q^2$. At some characteristic scale $Q^2 \approx Q_s^2(x)$,
the gluons in the proton start to overlap and so the linear
approximation is no longer applicable; non-linear corrections to the
linear evolution equations\cite{gribovnu} arise and become
significant, potentially taming the growth of the gluon 
distribution towards a  ``saturating'' behaviour.
%, i.e.
%towards a limitation of the maximum gluon density per unit of phase space.
\subsection{Instantons and Saturation?}
$eP$-scattering at small Bjorken-$x$ and
decreasing $Q^2$ uncovers a novel regime of QCD, where the coupling
$\alpha_s$ is (still) small, but the parton densities are so large
that conventional perturbation theory ceases to be applicable.  
Much interest has recently been generated through association of the
saturation phenomenon with a multiparticle quantum state of high
occupation numbers, the ``Colour Glass Condensate'' that  correspondingly,
can be viewed\cite{cgc} as a strong {\em classical} colour field
$\propto 1/\sqrt{\alpha_s}$. 

Being extended non-perturbative and topologically
non-trivial fluctuations of the gluon field, instantons\cite{bpst} ($I$) are
naturally very interesting in the context of saturation, since
\begin{itemize}
\item classical {\em non-perturbative} colour fields are physically appropriate;
\item the functional form of the instanton gauge fields is explicitely
      known and their strength is $A_\mu^{(I)}\propto
      \frac{1}{\sqrt{\alpha_s}}$ as needed;
\item an identification of the ``Colour Glass Condensate'' with the
      QCD-sphaleron state appears very suggestive\cite{su2,su3}
      (c.f. below and Sec~3.2).    
\end{itemize}
Two arguments in favour of instanton-driven
saturation are particularly worth emphasizing. 

We know already from $I$-perturbation theory that the instanton
contribution tends to strongly increase towards the softer
regime\cite{rs1,rs2,qcdins}. The mechanism for the decreasing
instanton suppression with increasing energy is known since a long
time\cite{sphal2,shuryak2}: Feeding increasing energy into the scattering
process makes the picture shift from one 
of tunneling between adjacent vacua ($E\approx 0$) to that of the actual
creation of the sphaleron-like, coherent multi-gluon 
configuration\cite{sphal1} on top of the potential barrier of
height\cite{rs1,diak-petrov} $E = m_{\rm sph}\propto\frac{1}{\alpha_s\rho_{\rm
eff.}}$.    
%%%%%%%%%%%%%%%%%%%%%%%%%%%%%%figure%%%%%%%%%%%%%%%%%%%%%%%%%%%%%%%
\begin{figure}[ht]
\centerline{\parbox{5.5cm}{\epsfxsize=5.5cm\epsfbox{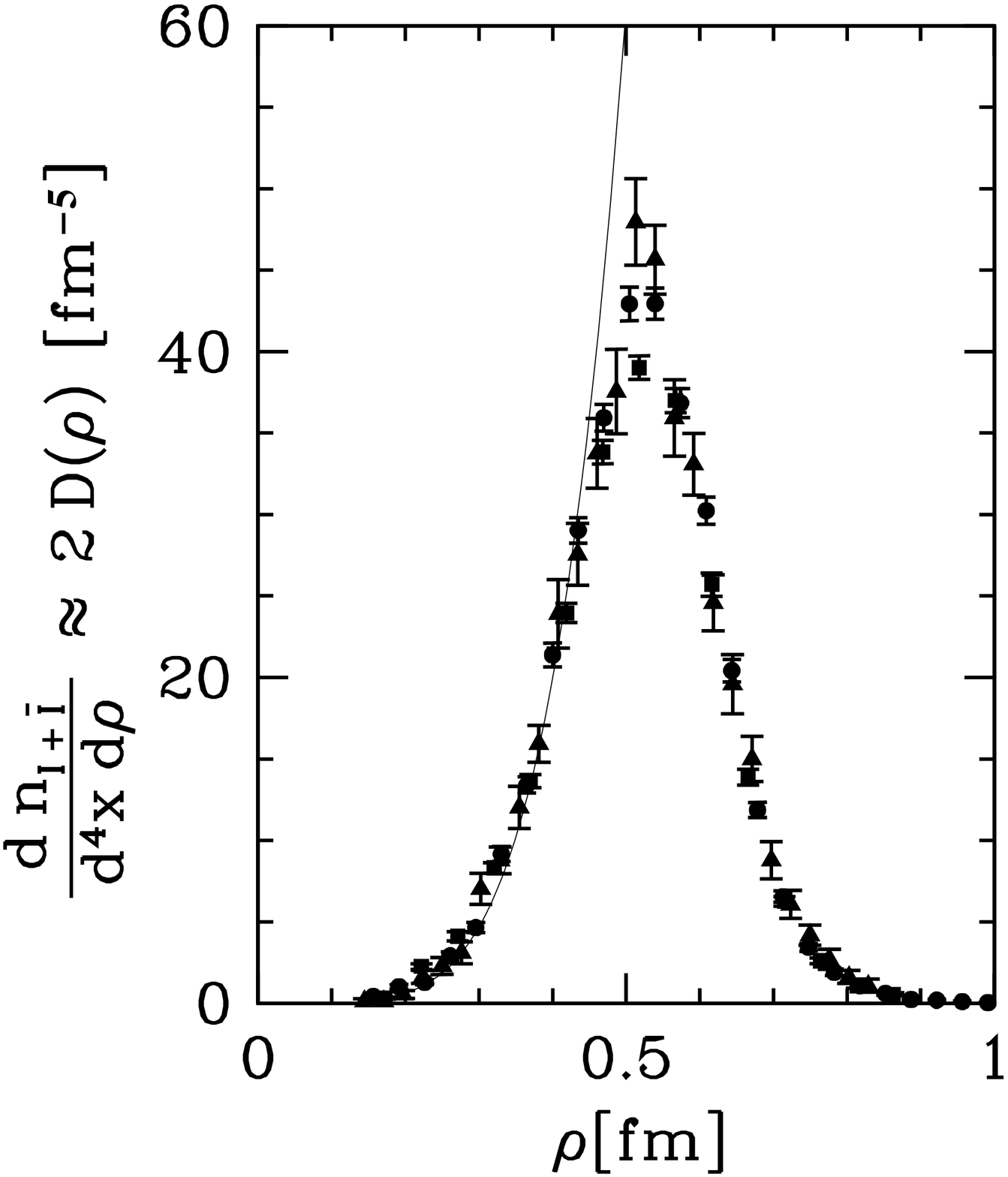}}   
\parbox{6.0cm}{\hspace{1.5cm}\parbox{4.0cm}{\epsfxsize=4.0cm
\epsfbox{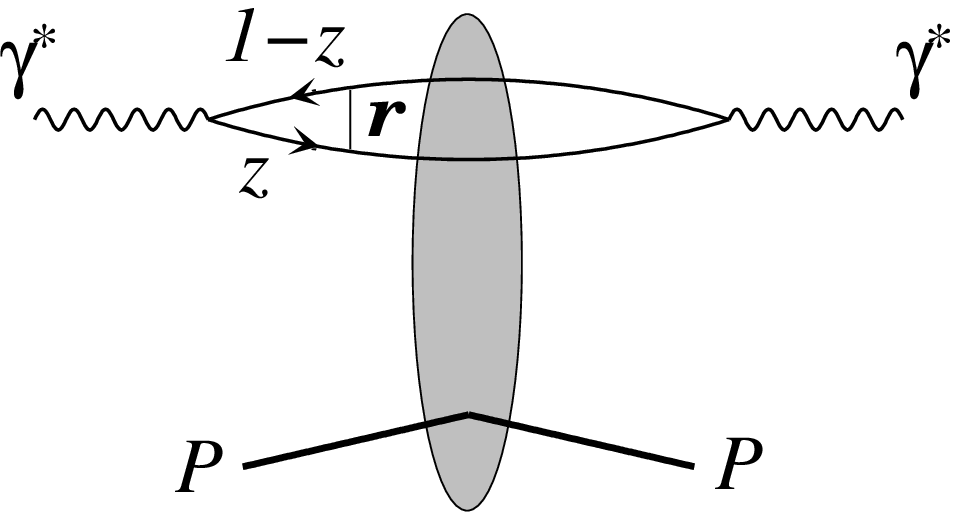}}
\parbox{6.0cm}{\vspace{0.25cm}\epsfxsize=6.0cm\epsfbox{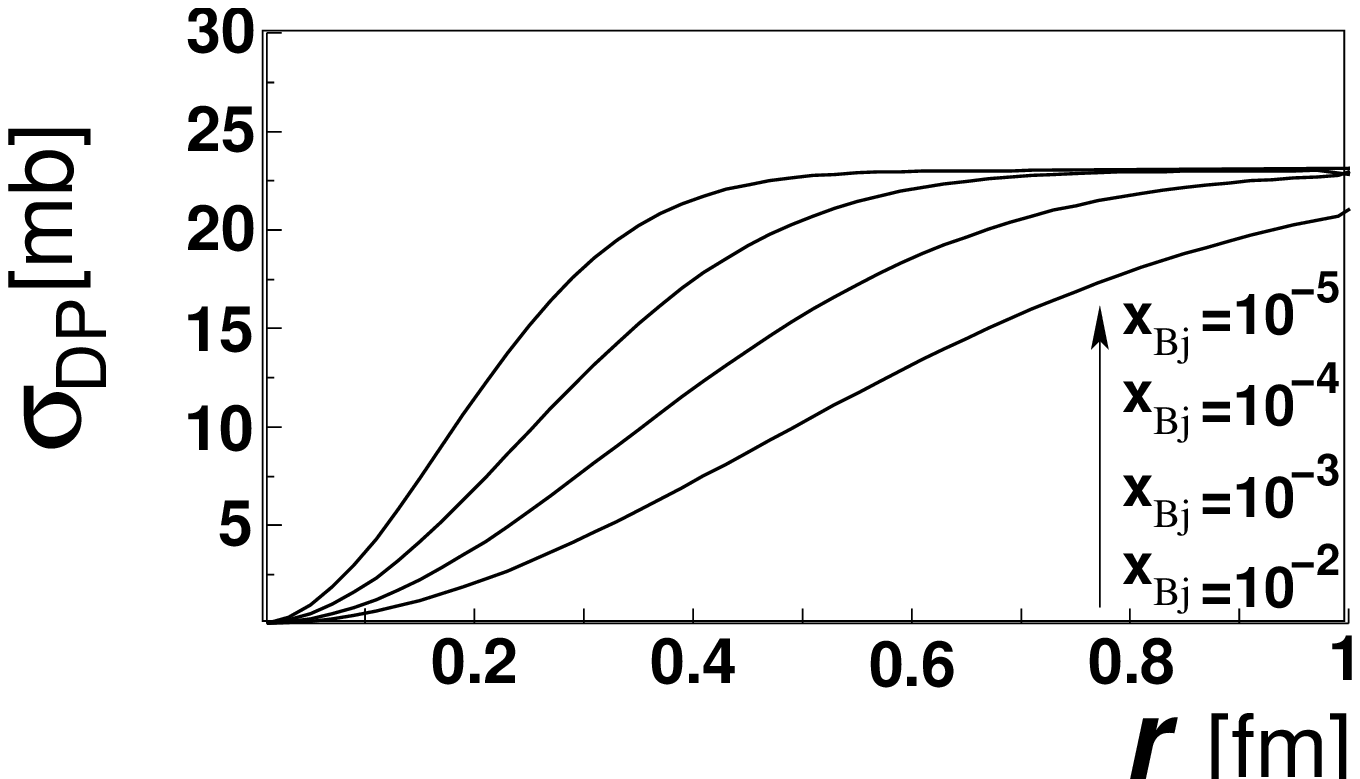}}}}
 \caption[dum]{(Left) UKQCD lattice data\cite{ukqcd,rs-lat,rs3} of the 
 $(I+\bar{I})$-size distribution for quenched QCD ($n_f = 0)$. Both
 the sharply defined $I$-size scale $\langle\rho\rangle \approx 0.5$
 fm and the parameter-free agreement with 
\mbox{$I$-perturbation} theory\cite{rs-lat,rs3} (solid line) for
$\rho\lwig 0.35$ fm are apparent.
(Right) $\gamma^\ast p$ scattering at small $x$. The photon
fluctuates into a $q\bar{q}$ dipole interacting with the proton (top).
Generic behaviour of the dipole-proton cross section, as taken from
the GBW-model\cite{gbw}(bottom). \label{pic1}}   
\end{figure}
%%%%%%%%%%%%%%%%%%%%%%%%%%%%%%%%%%%%%%%%%%%%%%%%%%%%%%%%%%%%%%%%%%%

A crucial aspect concerns the $I$-size $\rho$. On the one hand it is just a
collective coordinate to be integrated over in any observable, with
the $I$-size distribution $D(\rho)=d\,n_I/d^4z d\rho$ as universal weight. 
On the other hand, according to lattice data, $D(\rho)$ turns out to be {\em
sharply} peaked (Fig.~\ref{pic1} (left)) around $\rav\approx 0.5~{\rm fm}$.  
Hence instantons represent truly non-perturbative gluons that bring in
naturally an intrinsic size scale $\rav$ of hadronic dimension. As we
shall see, $\rav$ actually determines the saturation
scale\cite{su1,su2,su3}.   

Presumably, it is also reflected in the conspicuous {\it geometrization} of
soft QCD at high energies\cite{fs,su1,su2}. For related approaches
associating instantons with high-energy (diffractive) scattering, see
Refs.\cite{levin,shuryak1,shuryak11,shuryak2}. Instantons in the
context of small-$x$ saturation have also been studied recently by
Shuryak and Zahed\cite{shuryak3}, with conclusions differing in part
from those of our preceeding work\cite{fs,su1,su2,su3}. Their main
emphasis rests on Wilson loop scattering, and lattice information was
not used in their approach.  

In this paper we shall report about our results on the interesting 
possibility of instanton-driven saturation at small Bjorken-$x$. They
have been obtained by exploiting crucial non-perturbative information
from high-quality lattice simulations\cite{ukqcd,rs-lat}.

\section{Setting the Stage}

The investigation of saturation becomes most transparent in the
familiar colour-dipole picture\cite{dipole} (cf. Fig. 1 (top right)),
notably if analyzed in the so-called dipole frame\cite{mueller}. In this 
frame, most of the energy is still carried by the hadron, but the virtual
photon is sufficiently energetic, to dissociate before scattering into
a $q\bar{q}$-pair (a {\it colour dipole}), which then scatters off the
hadron. Since the latter is 
Lorentz-contracted, the dipole sees it as a colour source of 
transverse extent, living (essentially) on the light cone. This colour
field is created by the constituents of the well developed hadron wave
function and -- in view of its high intensity, i.e. large occupation
numbers -- can be considered as classical. Its strength near
saturation is $\mathcal{O}(1/\sqrt{\alpha_s})$. At high energies, 
the lifetime  of the $q\overline{q}$-dipole  is much larger than the
interaction time between this $q\overline{q}$-pair and the hadron and hence,
at small $\xbj$, this gives rise to the familiar factorized
expression of the inclusive photon-proton cross sections, 
\begin{equation}
\sigma_{L,T}(\xbj,Q^2) 
=\int_0^1 d z \int d^2\vr\; |\Psi_{L,T}(z,r)|^2\,\sigma_{\mbox{\tiny DP}}(r,\ldots).  
\label{dipole-cross}
\end{equation}
Here, $|\Psi_{L,T}(z,r)|^2$ denotes the modulus squared of the 
(light-cone) wave function of the virtual photon, calculable in pQCD,
and $\sigma_{\mbox{\tiny DP}}(r,\ldots)$ is the $q\overline{q}$-dipole\,-\,nucleon 
cross section.  The variables in Eq.~(\ref{dipole-cross}) are
the transverse $(q\overline{q})$-size $\mathbf r $ 
and the photon's longitudinal momentum fraction $z$ carried by the quark. 
The dipole cross section is expected to include in general the main
non-perturbative contributions. For small $r$, one finds within
pQCD\cite{dipole,dipole-pqcd} that $\sigma_{\mbox{\tiny DP}}$ vanishes 
with the area $\pi r^2$ of the $q\overline{q}$-dipole. Besides this phenomenon
of ``colour transparency'' for small $r=|\vr|$,  the dipole cross
section is expected to saturate towards a constant, once the
$q\overline{q}$-separation $r$ exceeds a certain saturation scale $r_s$. 
While there is no direct proof of the saturation phenomenon,
successful models incorporating saturation do exist\cite{gbw,gbwdglap}
and describe the data efficiently.

Let us outline more precisely the strategy we shall pursue:

The guiding question is: Can background instantons of size
$\sim\langle\rho\rangle$ give rise to a saturating, geometrical
form for the dipole cross section,
\begin{equation}
\sidp^{(I)}(r,\ldots)\stackrel{r\gwig \langle\rho\rangle}{\sim}\pi
      \langle\rho\rangle^2
\end{equation}
We have obtained answers from two alternative approaches\cite{fs,su1,su2,su3}:
\begin{enumerate}
\item {\em From $I$-perturbation theory to saturation:}
Here, we start from the large $Q^2$ regime and appropriate cuts
such that $I$-perturbation theory is strictly valid. The
corresponding, known results on $I$-induced DIS processes\cite{mrs}
are then transformed into the colour-dipole picture. With the crucial
help of lattice results, the $q\bar{q}$-dipole size $r$ is next
carefully increased towards hadronic dimensions. Thanks to the lattice
input, IR divergencies are removed and the original cuts are no longer
necessary. 
\item {\em Wilson-loop scattering in an $I$-background:}
As a second, complementary approach we have considered the
semi-classical, non-abelian eikonal approximation. It  results in 
the identification of the $q\bar{q}$-dipole with a Wilson loop,
scattering in the non-perturbative colour field of the proton. The field
$A_\mu^{(I)}\propto \frac{1}{\sqrt{\alpha_s}}$ 
due to background instantons is studied as a concrete example, leading
to analytically calculable results in qualitative agreement with the
first approach. 
\end{enumerate}

\section{From $I$-Perturbation Theory to Saturation}
\subsection{The Simplest Process: $\gamma^\ast+ g\stackrel{(I)}{\to} q_{\rm R}+\overline{q}_{\rm R}$}
%%%%%%%%%%%%%%%%%%%%%%%%%%%%%%figure%%%%%%%%%%%%%%%%%%%%%%%%%%%%%%%
\begin{figure}[ht]
\centerline{\parbox{10cm}{\epsfxsize=10cm
\epsfbox{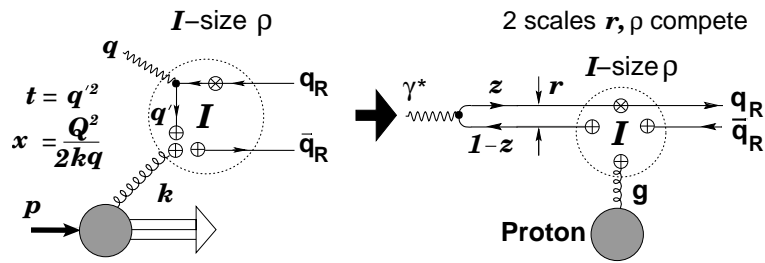}}}
 \caption[dum]{Transcription of the simplest $I$-induced process into
 the colour-dipole picture\label{simplest}}  
\end{figure}
%%%%%%%%%%%%%%%%%%%%%%%%%%%%%%%%%%%%%%%%%%%%%%%%%%%%%%%%%%%%%%%%%%%
Let us briefly consider first the simplest $I$-induced process, 
$\gamma^\ast\,g\Rightarrow q_{\rm R}\overline{q}_{\rm R}$, with one
flavour and no final-state gluons (Fig.~\ref{simplest} (left)). More details
may be found in Ref.\cite{su2}. Already this simplest case
illustrates transparently that in the presence of a background
instanton, the dipole cross section indeed saturates with a 
saturation scale of the order of the average $I$-size $\rav$.  

We start by recalling the results for the total $\gamma^\ast N$ cross
section within $I$-perturbation theory 
from Ref.\cite{mrs},  
\begin{eqnarray}
\sigma_{L,T}(\xbj,Q^2)&=&
\int\limits^1_{\xbj} \frac{d x}{x}\left(\frac{\xbj}{
x}\right)G\left(\frac{\xbj}{x},\mu^2\right)\int d  t \frac{d
\hat{\sigma}_{L,T}^{\gamma^* g}(x,t,Q^2)}{d t};\,\label{general}\\[2ex] 
\frac{d\hat{\sigma}_{L}^{\gamma^* g}}{d  t}&=&\frac{\pi^7}{2}
\frac{e_q^2}{Q^2}\frac{\alpha_{\rm em}}{\alpha_{\rm
s}}\left[x(1-x)\sqrt{t u}\,  \frac{\mathcal{R}(\sqrt{-
t})-\mathcal{R}(Q)}{t+Q^2}-(t\leftrightarrow  u)\right]^{\,2} \label{mrs}
\end{eqnarray}
with a similar expression for $d\hat{\sigma}_{T}^{\gamma^\ast\,g}/d\,t$. 
Here, $G\left(\xbj,\mu^2\right)$ denotes the gluon density and $L,T$
refers to longitudinal and transverse photons, respectively.

Note that Eqs.~(\ref{general}),~(\ref{mrs}) involve the ``master'' integral
$\mathcal{R}(\mathcal{Q})$ with dimension of a length,  
\begin{equation}
\mathcal{R}(\mathcal{Q})=\int_0^{\infty} d\rho\;D(\rho)\rho^5(\mathcal{Q}\rho)\mbox{K}_1(\mathcal{Q}\rho).
\label{masterI}
\end{equation}
In usual $I$-perturbation theory, the $\rho$-dependence of the $I$-size
distribution $D(\rho)$ in Eq.(\ref{masterI}) is known\cite{th} for
sufficiently small $\rho$,  
\begin{equation}
D(\rho)\approx D_{I-{\rm pert}}(\rho) \propto \rho^{6-\frac{2}{3}n_f},
\end{equation}
the strong power law increase of which, is well known to generically lead to
(unphysical) IR-divergencies from large-size instantons. However, in DIS for
sufficiently large virtualities $\mathcal{Q}$, the crucial factor
$(\mathcal{Q}\rho)\,K_1(\mathcal{Q}\rho)\sim e^{-\mathcal{Q}\rho}$
in Eq.(\ref{masterI}) exponentially suppresses large size instantons
and $I$-perturbation theory holds, as shown first in Ref.\cite{mrs}.

The effective size $\mathcal{R}(\mathcal{Q})$ in Eq.(\ref{masterI})
correspondingly plays a central r\^{ole} in the context of a
continuation of our $I$-perturbative results to smaller
$\mathcal{Q}$. Here, crucial lattice information enters. We recall
that the  $I$-size distribution $D_{\rm lattice}(\rho)$, as {\it measured} on the lattice\cite{ukqcd,rs-lat,rs3}, is strongly peaked around an average
$I$-size $\langle\rho\rangle \approx 0.5$ fm (cf. Fig.~\ref{pic1}
(left)), while being in excellent, 
parameter-free agreement\cite{rs-lat,rs3} with $I$-perturbation
theory for $\rho\lwig 0.35$ fm (cf. Fig.~\ref{pic1} (left)).   

Our general strategy is thus to generally identify $D(\rho) = D_{\rm
lattice}(\rho)$ in Eq.(\ref{masterI}), whence 
\begin{equation}
\mathcal{R}(0)=\int_0^{\infty} d\rho\;D_{\rm
lattice}(\rho)\rho^5\approx 0.3 \mbox{\ fm}
\end{equation} 
becomes finite and a $\mathcal{Q}^2$ cut is no longer necessary. 

By means of an appropriate change of variables 
and a subsequent $2d$-Fourier transformation,
Eqs.~(\ref{general}), (\ref{mrs}) may 
indeed be cast\cite{su2} into a colour-dipole form
(\ref{dipole-cross}), e.g. (with $\hat{Q}=\sqrt{z\,(1-z)}\,Q$)
\begin{eqnarray}
\lefteqn{\left(\left|\Psi_L\right|^2\sigma_{\mbox{\tiny DP}}\right)^{(I)}
 \approx\, \mid\Psi_L^{\rm pQCD}(z,r)\mid^{\,2}\,
\frac{1}{\alpha_{\rm s}}\,\xbj\,
G(\xbj,\mu^2)\,\frac{\pi^8}{12}}\label{resultL}\\[1ex] 
&&\times\left\{\int_0^\infty\,d\rho
D(\rho)\,\rho^5\,\left(\frac{-\frac{d}{dr^2}\left(2 r^2 
\frac{\mbox{K}_1(\hat{Q}\sqrt{r^2+\rho^2/z})}{\hat{Q}\sqrt{r^2+\rho^2/z}}
\right)}{{\rm K}_0(\hat{Q}r)}-(z\leftrightarrow 1-z)
\right)\right\}^2.\nonumber 
\end{eqnarray} 
The strong peaking of $D_{\rm
lattice}(\rho)$ around \mbox{$\rho\approx\rav$}, implies 
\begin{equation}
\left(|\Psi_{L,T}|^{\,2}
\sigma_{\mbox{\tiny DP}}\right)^{(I)}\Rightarrow\left\{\begin{array}{llcl} 
\mathcal{O}(1) \mbox{\rm \ but exponentially small};&r\rightarrow 0,\\[2ex]
\mid\Psi^{\rm \,pQCD}_{L,T}\mid^{\,2}\,\frac{1}{\alpha_{\rm
s}}\,\xbj\,G(\xbj,\mu^2)\,\frac{\pi^8}{12}\,\mathcal{R}(0)^2;
&\frac{r}{\rav}\gwig 1.\label{final} 
\end{array}\right.
\end{equation} 
Hence, the association of the intrinsic instanton scale $\rav$ with
the saturation scale $r_s$ becomes apparent from Eqs.~(\ref{resultL}),
(\ref{final}): $\sigma_{\mbox{\tiny DP}}^{(I)}(r,\ldots)$ rises strongly as function of
$r$ around $r_s\approx\langle\rho\rangle$, and  
indeed {\em saturates} for $r/\rav>1$  towards a {\em constant
geometrical limit}, proportional to the area
$\pi\,\mathcal{R}(0)^2\, =\,  \pi\left(\int_0^\infty\,d\rho\,D_{\rm
lattice}(\rho)\,\rho^5\right)^2$, subtended by the instanton.
Since $\mathcal{R}(0)$ would be divergent within
$I$-perturbation theory, the information about $D(\rho)$ from the 
lattice (Fig.~\ref{pic1} (left)) is crucial for the finiteness of the result. 

\subsection{The Realistic Process: $\bm{\gamma^\ast + g\stackrel{(I)}{\to}
n_f\,(q_{\rm R} +\overline{q}_{\rm R})} + \mbox{gluons}$}
On the one hand, the inclusion of an arbitrary number of final-state
gluons and $n_f>1$ light flavours causes a significant complication. On
the other hand, it is due to the inclusion of final-state gluons that the 
identification of the QCD-sphaleron state with the colour glass
condensate has emerged\cite{su2,su3}, with the qualitative ``saturation''
features of the preceeding subsection remaining unchanged. In view of
the limited space, let us therefore focus our main  attention in this
section to the emerging sphaleron interpretation of the colour glass
condensate. 

Most of the $I$-dynamics resides in $I$-induced $q^\ast
g$-subprocesses like 
\begin{equation}
q_{\rm L}^\ast( q^\prime)+g(p)\xrightarrow{(I)} n_f\,q_{\rm R}
+(n_f-1)\,\bar{q}_{\rm R}+{\rm gluons}, 
\end{equation}
with an incoming off-mass-shell quark $q^\ast$ originating from photon
dissociation, $\gamma\to\bar{q}+q^\ast$. The important kinematical
variables are the total $I$-subprocess energy
$E=\sqrt{(q^\prime+p)^2}$ and the quark 
virtuality $Q^{\prime\,2}=-q^{\prime\,2}$. 

%%%%%%%%%%%%%%%%%%%%%%%%%%%%%%figure%%%%%%%%%%%%%%%%%%%%%%%%%%%%%%%
\begin{figure}[t]
\centering
\parbox{11.5cm}{\epsfxsize=11.5cm\epsfbox{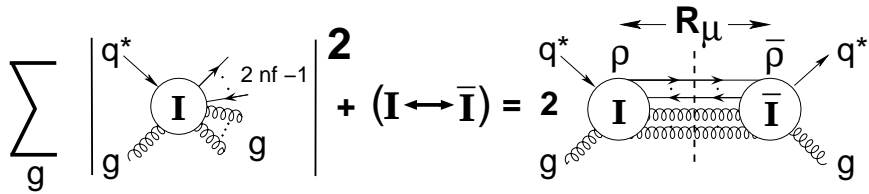}}
 \caption[dum]{Optical theorem for the $I$-induced $q^\ast g$-
 subprocess. The incoming, virtual $q^\ast$ originates from photon
 dissociation, $\gamma\to\bar{q}+q^\ast$.  \label{optfig}}  
\end{figure}
%%%%%%%%%%%%%%%%%%%%%%%%%%%%%%%%%%%%%%%%%%%%%%%%%%%%%%%%%%%%%%%%%%%
It is most convenient to account for the arbitrarily many final-state
gluons by means of the so-called ``$I\bar{I}$-valley
method''\cite{yung}. It allows to achieve via the optical theorem
(cf. Fig.~\ref{optfig}) an elegant summation over the final-state
gluons in form of an exponentiation, with the  
effect of the gluons residing entirely in the $I\bar{I}$-valley interaction
\mbox{$-1\le\Omega_\text{valley}^{I\bar{I}}(\frac{R^2}{\rho\bar{\rho}}+\frac{\rho}{\bar{\rho}}+\frac{\bar{\rho}}{\rho};U)
\le 0$}, between $I$'s and $\bar{I}$'s. The new collective coordinate
$R_\mu$ denotes the $I\bar{I}$-distance 4-vector
(cf. Fig.~\ref{optfig}), while the matrix $U$ characterizes the
$I\bar{I}$ relative colour orientation. Most
importantly, the functional form of $\Omega_\text{valley}^{I\bar{I}}$
is analytically known\cite{kr,verbaarschot} and the limit of $I$-perturbation
theory is attained for $\sqrt{{R^{\,2}}}\gg \sqrt{{ \rho\bar{\rho}}}$.     

The strategy we shall apply, is identical to the one for the ``simplest
process'' in the previous Sec.~3.1: Starting point is the general form
of the $I$-induced $\gamma^\ast N$ cross section, this time obtained by means of the  $I\bar{I}$-valley method\cite{rs2}. By exploiting the optical theorem
(cf. Fig.~\ref{optfig}), the total $q^\ast g$-cross
section is most efficiently evaluated from the imaginary part of the
forward elastic 
amplitude induced by the $I\bar{I}$-valley background. The next step is
again a variable and $2d$-Fourier transformation into the colour-dipole
picture like before. 

The dipole cross section $\tilde{\sigma}^{(I),{\rm
    gluons}}_{\mbox{\tiny DP}}(\vl^2,\xbj,\ldots)$ before the final
$2d$-Fourier transformation\footnote{The
  2-dimensional vector $\vl$ denotes the transverse momentum of the
  quark with four-momentum  $q^\prime$} $\vl\leftrightarrow\vr$ to the dipole size
$\vr$, arises simply as an energy integral over the $I$-induced total
$q^\ast g$ cross section from Ref.\cite{rs2},
\begin{equation}
\tilde{\sigma}^{(I),{\rm gluons}}_{\mbox{\tiny DP}} \approx
\frac{\xbj}{2}\,G(\xbj,\mu^2)\,\int_0^{E_{\rm max}}
\frac{d\,E}{E} \left[\frac{E^4}{(E^2+Q^{\,\prime 2})\,Q^{\,\prime 2}}\, 
\sigma^{(I)}_{q^\ast \,g}\left(E,\vl^2,\ldots\right)\right],
\label{sigdipglue}
\end{equation}
involving in turn integrations over the $I\bar{I}$-collective coordinates
$\rho,\bar{\rho},U$ and the $I\bar{I}$-distance $R_\mu$.

\begin{figure}[t]
\centerline{\parbox{5.2cm}{\epsfxsize=5.2cm
\epsfbox{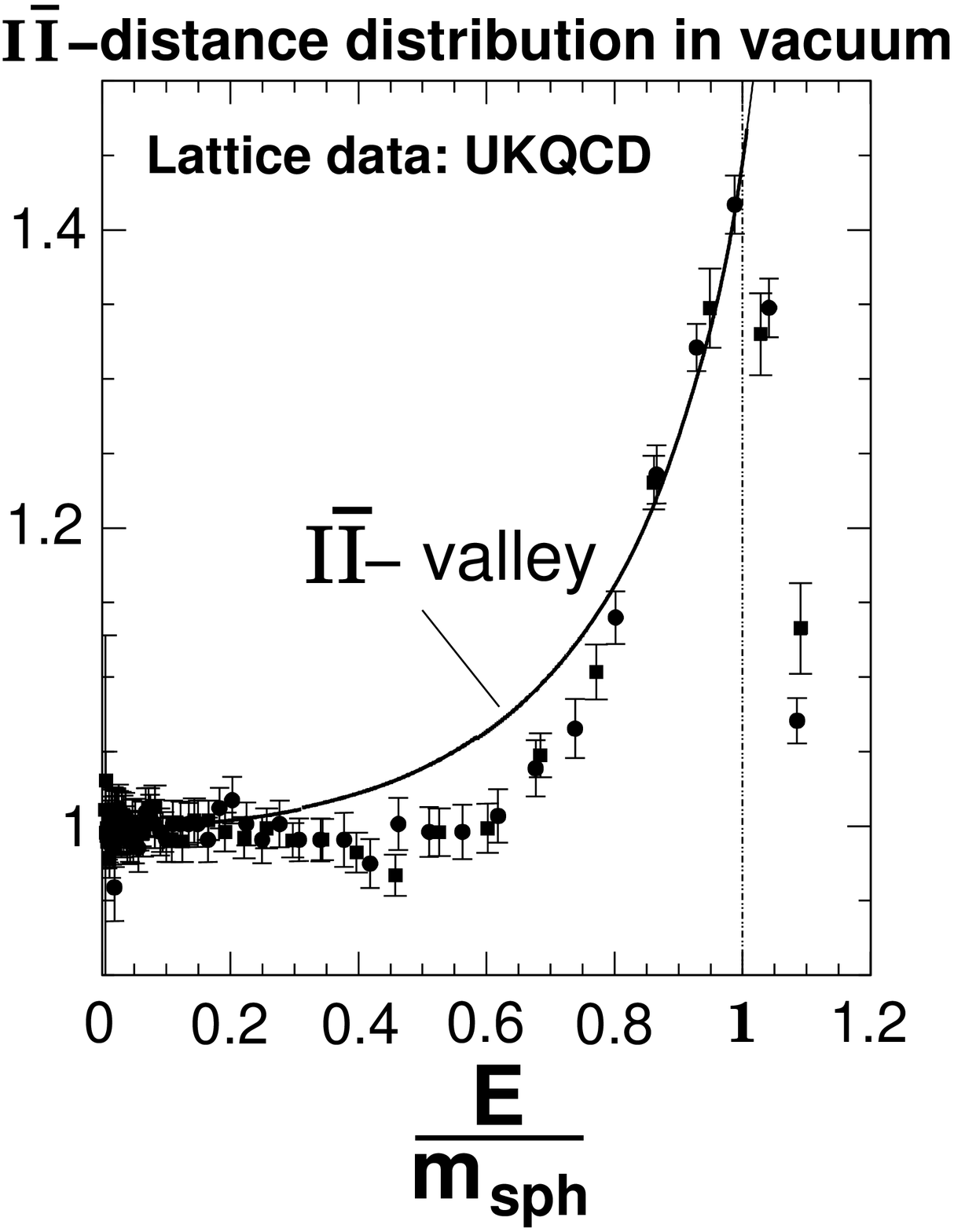}}\hfill   
\parbox{5.8cm}{\vspace{0.2cm}\epsfxsize=5.8cm\epsfbox{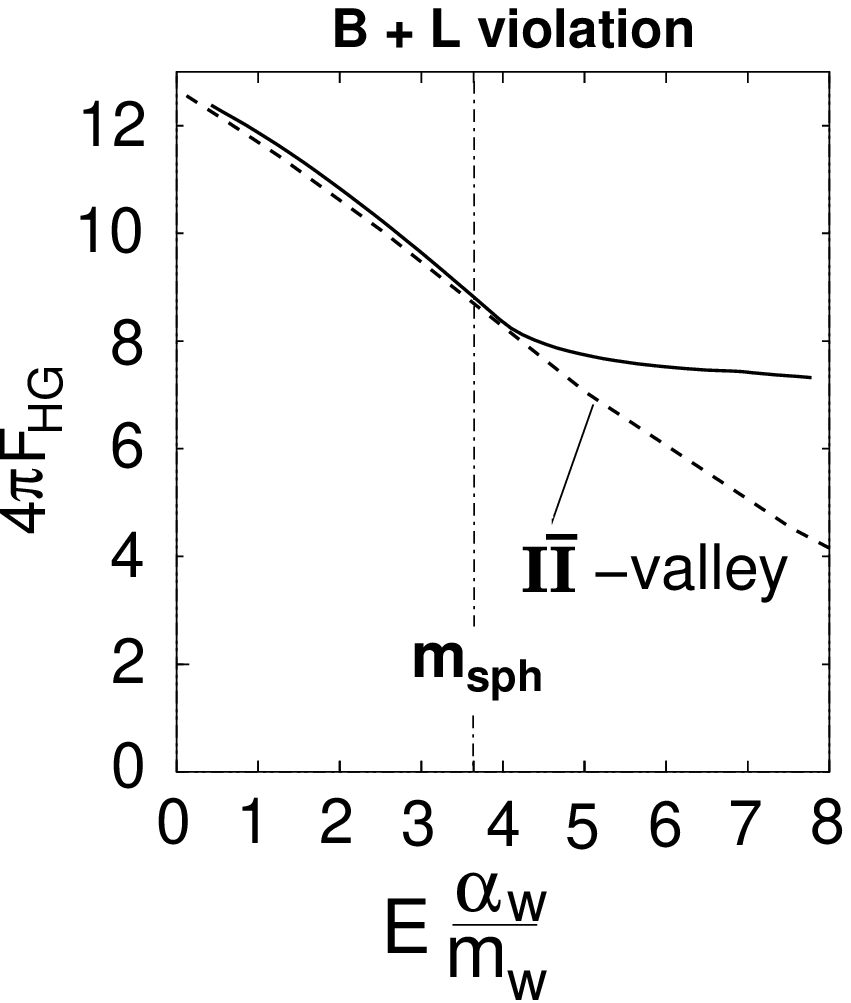}}}
 \caption[dum]{(Left) UKQCD lattice
 data\cite{ukqcd,rs-lat}  of the (normalized) $I\bar{I}$-distance
 distribution and the corresponding $I\bar{I}$-valley
 prediction\cite{su2} displayed versus energy in units of the QCD
 sphaleron mass $m_{\rm sph}$. The lattice data provide the first direct
 evidence that the $I\bar{I}$-valley approach is adequate
 right up to $E\approx m_{\rm sph}$, where the dominant contribution to the
 scattering process arises. Beyond this point a marked disagreement
rapidly develops. (Right) The same trend for electroweak $B
 +L$-violation is apparent from a completely independent
 semiclassical, numerical simulation of the suppression exponent for
 two-particle collisions ('Holy Grail' function) $F_{\rm
 HG}(E)$\cite{rubakov,ringwald}.    
 \label{pic2}}  
\end{figure}

In the softer regime of interest for saturation, we again substitute
$D(\rho) = D_{\rm lattice}(\rho)$, which enforces $\rho\approx\bar{\rho}\approx
\langle\rho\rangle$ in the respective $\rho,\bar{\rho}$-integrals,
while the integral over the $I\bar{I}$-distance $R$ is dominated by a
{\it saddle point}, 
\begin{equation}
\frac{R}{\rav} \approx {\rm
function}\left(\frac{E}{m_{\rm sph}}\right); \hspace{2ex} m_{\rm
sph}\approx \frac{3\pi}{4}\frac{1}{\alpha_s\,\rav} =\mathcal{O}({\rm
\,few\ GeV\,}).
\label{sphaleron1}
\end{equation}
At this point, the mass $m_{\rm sph}$ of the
QCD-sphaleron\cite{rs1,diak-petrov}, i.e the barrier height
separating neighboring topologically inequivalent vacua, enters as the
scale for the $I$-subprocess energy $E$. The saddle-point dominance of
the $R$-integration implies a one-to-one relation, 
\begin{equation} 
\frac{R}{\rav} \Leftrightarrow \frac{E}{m_{\rm sph}}; \hspace{2ex}
\mbox{\rm with}\ R=\rav \Leftrightarrow E\approx m_{\rm sph}.
\label{sphaleron2}
\end{equation}
Our careful continuation to the saturation regime now involves
in addition to the $I$-size distribution $D_{\rm lattice}(\rho)$, crucial
lattice information about the second basic building block 
of the $I$-calculus, the $I\bar{I}$-interaction $\Omega^{I\bar{I}}$. 
The relevant lattice observable is the distribution of the
$I\bar{I}$-distance\cite{rs-lat,su2} $R=\sqrt{R_\mu^2}$, 
essentially  providing information on $\left\langle\exp
[-\frac{4\pi}{\alpha_s}\Omega^{I\bar{I}}]\right\rangle_{U,\rho,\bar{\rho}}$
in Euclidean space. Due to the crucial saddle-point
relation~(\ref{sphaleron1}),~(\ref{sphaleron2}), 
we may replace the original variable $R/\rav$ by $E/m_{\rm sph}$. 
A comparison of the respective $I\bar{I}$-valley predictions with the
UKQCD lattice data\cite{ukqcd,rs-lat,su2} versus $E/m_{\rm sph}$ is
displayed in Fig.~\ref{pic2} (left). It reveals the important result
that the $I\bar{I}$-valley approximation is quite reliable up to
$E\approx m_{\rm sph}$. Beyond this point a marked disagreement
rapidly develops: While the lattice data show a {\it sharp peak} at
$E\approx m_{\rm sph}$, the valley prediction continues to rise
indefinitely for $E\gwig m_{\rm sph}$! It is most remarkable that an
extensive recent and completely independent semiclassical numerical
simulation\cite{rubakov} shows
precisely the same trend for electroweak $B +L$-violation, as
displayed in Fig.~\ref{pic2} (right). Also here, there is an amazing
agreement with the valley approximation\cite{ringwald} up to the electroweak
sphaleron mass, and a rapid disagreement developing beyond. 
%%%%%%%%%%%%%%%%%%%%%%%%%%%%%%figure%%%%%%%%%%%%%%%%%%%%%%%%%%%%%%%
%%%%%%%%%%%%%%%%%%%%%%%%%%%%%%%%%%%%%%%%%%%%%%%%%%%%%%%%%%%%%%%%%%%
It is again at hand to identify $\Omega^{I\bar{I}}=
\Omega^{I\bar{I}}_{\rm lattice}$ for $E\gwig m_{\rm sph}$.
Then, on account of Eq. (\ref{sigdipglue}), the integral over the
$I$-subprocess energy spectrum in the dipole cross section appears to 
be dominated by the sphaleron configuration at $E\approx m_{\rm sph}$.
The feature of saturation analogously to the ``simplest process'' in
Sec.~3.1 then implies the announced identification of the colour glass
condensate with the QCD-sphaleron state.
\section{Wilson-Loop Scattering in an $I$-Background}
Let us next turn to our second approach within the
colour-dipole picture, which is still in progress\cite{su3,su4}. It
is complementary to our previous strategy of 
extending the known results of $I$-perturbation theory towards the
saturation regime by means of non-perturbative lattice information. 
$q\bar{q}$-dipole scattering will be described as the scattering of
Wilson loops. 

We work within  the semiclassical, non-abelian eikonal
approximation that is appropriate for the scattering of partons at
high energies ($s\gg-t$ or small $\xbj$) from a soft colour field
$A_\mu$\cite{eik}. The basic approximation is that the soft
interaction of the partons with the colour field does not change their
direction appreciably, such that they just pick up a non-abelian phase
factor during the scattering. Each phase factor is given by a
path-ordered integral calculated along the classical path of the
respective parton, 
\begin{equation}
 W(\vx)={\rm P}\,\exp\left\{
-i\,g_s\, \int_{- \infty}^{+\infty} d\,\lambda\; q^\mu\,
A_\mu(\lambda\,q+{x_\perp})\right\}\,,\ {\rm with}\ x_\perp\cdot q=0.
\end{equation}
This so-called Wilson line, depends on the 2-dimensional
vector $\vx$ describing the distance to the proton-photon plane.
Correspondingly, $q\bar{q}$-dipoles lead to colourless, gauge invariant
Wilson loops: 
\begin{equation}
\mathcal{W}(\vr,\vb;A_\mu)=\frac{1}{N_c}\,{\rm tr}\left[W\left(\vb+\vr
    /2\right)
W^\dagger\left(\vb-\vr/2\right)\right]\,.
\label{WL}
\end{equation}
The Wilson loop (\ref{WL}) depends on the transverse size $\vr$ of the
colour dipole and the transverse distance $\vb$ between dipole
and colour field $A_\mu$. It is a basic object in the framework 
of the colour glass condensate language, where the proton is viewed as a
source of the classical field $A_\mu$. Averaging over possible field
configurations ($\langle\ldots\rangle_{A_\mu}$) and integrating over
the impact parameter $\vb$ leads to the total dipole cross
section (e.g.\cite{balitsky,cgc}),
\begin{equation}
\sigma_{\mbox{\tiny DP}}(r,\ldots)=2\,\int d^2\vb\left\langle
  1-\mathcal{W}(\vr,\vb;A_\mu)\right\rangle_{A_\mu}\,.
\label{sidp}
\end{equation}
In general, the meaning of the colour glass condensate in the context
of the dipole cross section (\ref{sidp}) (cf. also
Ref.\cite{buchmuwi}) is that of an effective 
theory, leading to non-linear evolution equations\cite{balitsky,K} for
the respective scattering amplitude. 

As a first concrete testing ground for the impact of instantons within
this framework, let us identify the classical
field $A_\mu$ in the dipole cross section (\ref{sidp}) with the known
instanton field $A^{(I)}_\mu$. The functional integration
$\langle\ldots\rangle_{A_\mu}$ over the field configurations $A_\mu$
is then to be understood as an integration over the I-collective
coordinates, i.e. 
\begin{equation}
A_\mu(x)\to A^{(I)}_\mu (x,\rho,x_0);\hspace{2ex}
 \langle\ldots\rangle_{A_\mu}\to \mathcal{D}A^{(I)}_\mu \to
 d^4x_0\,d\rho\,D_{\rm  lattice}(\rho)\,.
\end{equation}
In a  first step, one has to calculate the Wilson loop in the 
$I$-background, which can be performed analytically. Subsequently, one
has to integrate over the collective coordinates. Finally, we get a
dipole cross section depending on the dipole size $r$ and the size
$\langle\rho\rangle$ of the instanton in the vacuum,
\begin{equation}
\sigma_{\mbox{\tiny
DP}}^{(I)}(r,\ldots)\propto{\langle\rho\rangle^2}\,f
\left(\frac{r}{\langle\rho\rangle}\right)\,.
\end{equation}
Like in our first approach (Sec. 3), this dipole cross section turns
out to saturate
towards a constant limit proportional to $\langle\rho\rangle^2$ for
$r\gwig\rav$.   

\begin{figure}[ht]
\centerline{\parbox{4.0cm}{\epsfxsize=4.0cm
\epsfbox{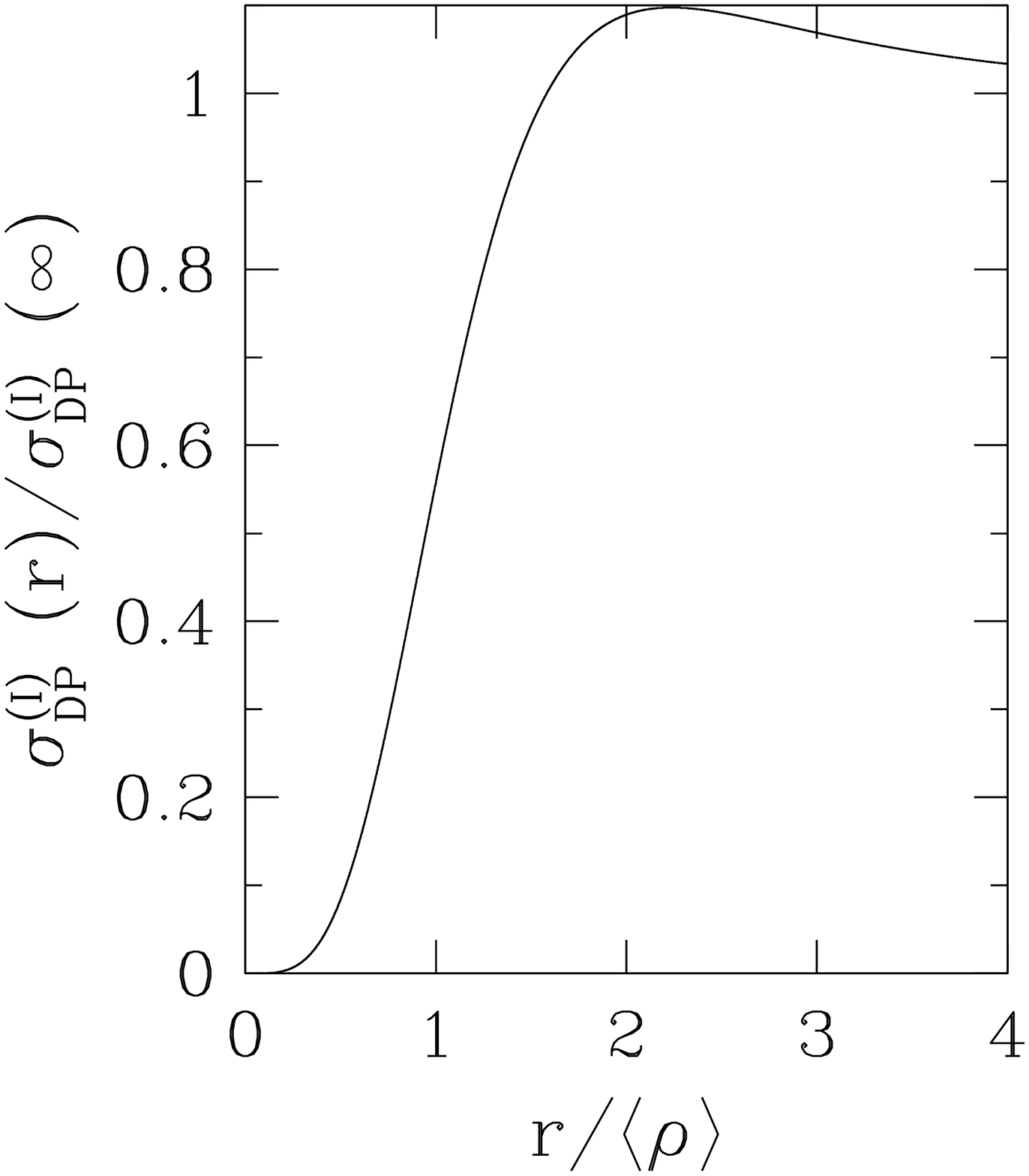}}\hspace{1cm}   
\parbox{4.0cm}{\epsfxsize=4.0cm\epsfbox{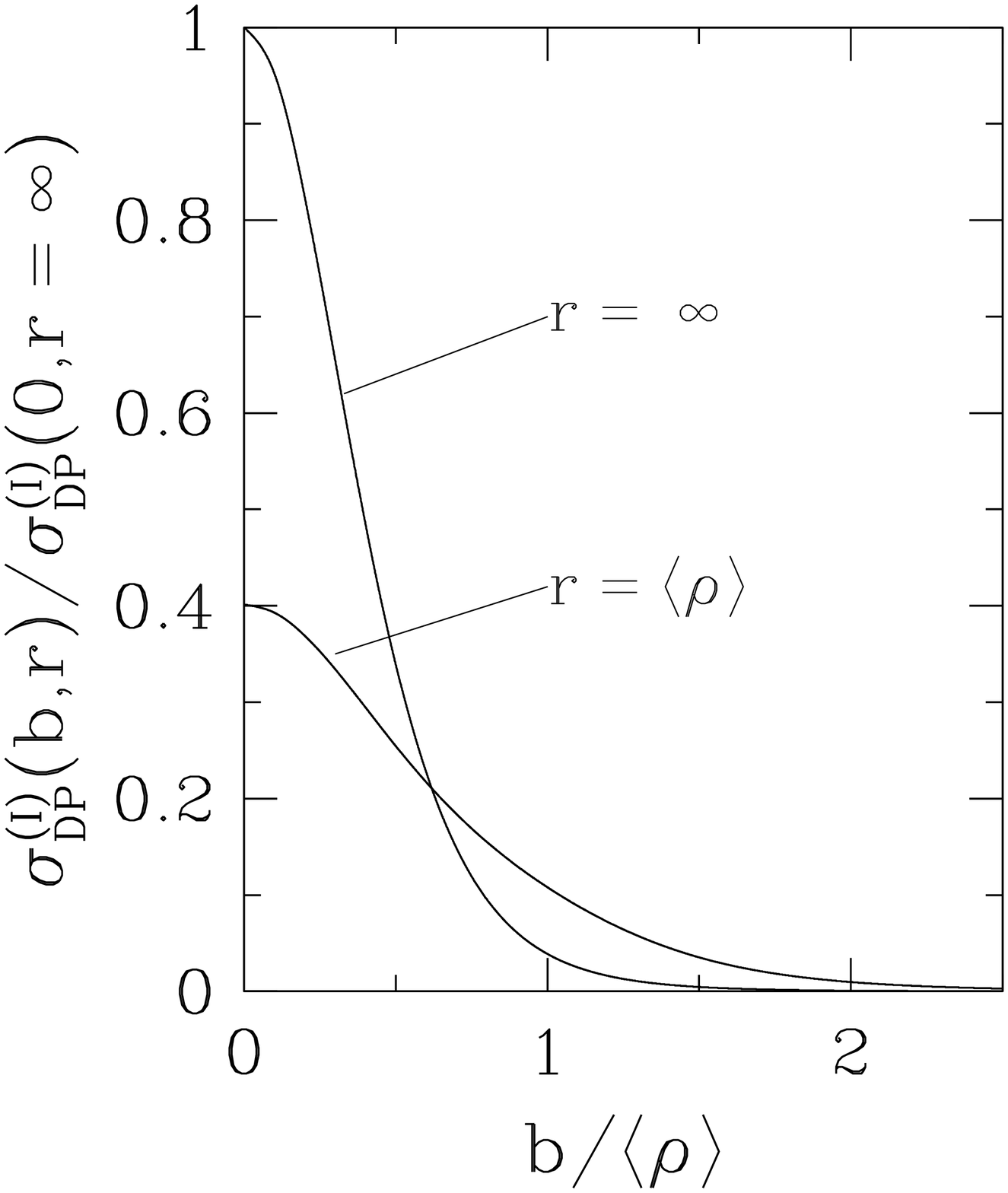}}}
\caption[dum]{(Left) The elastic contribution to the I-induced dipole cross
section \\
(Right) The corresponding impact parameter profile. \label{pic3}}  
\end{figure}

For a more realistic estimate,  it is important to notice that one has
to take an 
$I\bar{I}$-configuration (like the valley field\cite{su4}) in the total cross
section (\ref{sidp}). For an estimate of the elastic part of the
dipole cross section, one can take the single $I$-gauge field and
square the resulting dipole scattering-amplitude. This elastic
contribution $\sigma^{(I)}_{\mbox{\tiny 
DP}}(r)/\sigma^{(I)}_{\mbox{\tiny DP}}(\infty)$, normalized to one
for $r\to\infty$, is displayed in Fig.~\ref{pic3} (left) as function of
$r/\rav$. The importance of $\rav$ in the approach to
saturation becomes again apparent. In Fig.~\ref{pic3} (right), the
corresponding impact parameter profile for $r=\rav,\ \infty$ is displayed.
%%%%%%%%%%%%%%%%%%%%%%%%%%%%%%figure%%%%%%%%%%%%%%%%%%%%%%%%%%%%%%%
%%%%%%%%%%%%%%%%%%%%%%%%%%%%%%%%%%%%%%%%%%%%%%%%%%%%%%%%%%%%%%%%%%%
This simplest estimate of the dipole cross
section in an $I$-background can certainly not describe the proton in
an adequate way, notably due to the lack of non-trivial proton
kinematics. Hence it is not surprising  that the resulting dipole
cross section and hence the saturation scale comes out   
$\xbj$-independent in this case. Nevertheless, this
calculation illustrates once more the close connection between an 'extended'
classical colour background field of size $\rav$
and the saturation scale. Taking the instanton solution as an initial
condition for the BK-equations\cite{balitsky,K}, one could generate the
proper $\xbj$-dependence via the implied evolution. 

In order to model the proton more realistically, we have 
also worked out\cite{su4} a generalization to dipole-dipole
scattering in an $I$-background. The formalism used is analogous to  
Ref.\cite{shoshi} within the stochastic vacuum approach\cite{dosch}.
Like in Ref.\cite{shoshi}, we started
with the calculation of the loop-loop contributions to the
dipole-dipole scattering-amplitude, that are dominant in the
large-$N_c$ limit. In this case, the trace in Eq.~(\ref{WL}) is
taken separately for both dipoles. This leads to consistent
results and we observes again a saturation of the resulting dipole
cross section. In Refs.\cite{shuryak1,shuryak11,shuryak3} it was
pointed out, however, that the dominant contribution in the high-energy limit
comes from contribution with a colour exchange between the
dipoles. The impact of this contribution for the dipole cross section
is presently under investigation\cite{su4}. 
\section*{Acknowledgments}
We are grateful to Andreas Ringwald for a careful reading of the
manuscript.

\end{document}